\documentclass[twocolumn,pra,floatfix,citeautoscript,nofootinbib,superscriptaddress]{revtex4}
\usepackage{amsbsy}
\usepackage{latexsym,epsfig,graphicx}
\usepackage{dcolumn}
\usepackage{graphicx}
\usepackage{subfigure}
\usepackage{comment}
\usepackage{color}
\usepackage{bm}
\usepackage{mathrsfs}
\usepackage{amsfonts}
\usepackage{amsmath}
\usepackage{color}
\usepackage{amssymb}
\usepackage{xspace}
\usepackage{epstopdf}
\usepackage{tabularx}
\usepackage{longtable}
\usepackage[colorlinks=true, letterpaper=true, pdfstartview=FitV, linkcolor=blue, citecolor=blue, urlcolor=blue]{hyperref}
\usepackage[normalem]{ulem}

\pdfoutput=1
\input{tcilatex}
\begin{document}

\title{Cancellation theorem breaking and resonant spin-tensor Hall
conductivity in higher-rank spin-tensor Hall effects}
\author{Xiaoru He}
\affiliation{School of Sciences, Xi'an Technological University, Xi'an 710032, China}
\author{Ling-Zheng Meng}
\affiliation{School of Sciences, Xi'an Technological University, Xi'an 710032, China}
\author{Junpeng Hou}
\email{jhou@pinterest.com}
\affiliation{Pinterest Inc., San Francisco, California 94103, United States}
\author{Xi-Wang Luo}
\email{luoxw@ustc.edu.cn}
\affiliation{CAS Key Laboratory of Quantum Information, University of Science and
Technology of China, Hefei, Anhui 230026, China}
\author{Ya-Jie Wu}
\email{wuyajie@xatu.edu.cn}
\affiliation{School of Sciences, Xi'an Technological University, Xi'an 710032, China}

\begin{abstract}
With recent advances in simulating quantum phenomena in cold atoms, the
higher-rank spin tensor Hall effect was discovered in larger spin systems
with spin-tensor-momentum coupling, which is an extension of the celebrated
spin Hall effects in larger spins. Previously, it has been proposed that a
2D electron gas with Rashba spin-orbit coupling can generate dissipationless
transverse spin current, namely the spin Hall effect. However, later work
showed that the spin current is canceled by vertex correction, which was
subsequently proven by a cancellation theorem that does not depend on any
assumptions related to the scattering mechanism, the strength of spin-orbit
coupling, or the Fermi energy. While the recent proposal demonstrates a
universal intrinsic spin-tensor Hall conductivity, it is unclear if it
vanishes similarly to the spin Hall effect. In this work, we address this
critical problem and show that the rank-2 spin-tensor current can be
divergent by considering the contributions of both interbranch and
intrabranch transitions, which resembles the quantum Hall effect in some
sense. So the \textit{universal} spin-tensor Hall effect can not be observed
in a system with finite size. However, we further show that there is an
\textit{observable non-zero} resonance of spin-tensor Hall conductivity as
the Landau levels cross under the magnetic field. Our work reveals
interesting conductivity properties of larger-spin systems and will provide
valuable guidance for experimental explorations of higher-rank spin-tensor
Hall effects, as well as their potential device applications.
\end{abstract}

\maketitle






\section{Introduction}

The discovery of Hall effects and spin Hall effects opens avenues for
generating dissipationless transverse charge and spin current driven by DC
electric field \cite{1,2,3}. The spin Hall effect provides a novel way to
control spins and implement low-power spintronic devices without the help of
strong magnetic fields in contrast to the Hall effect. Thus it has attracted
great interest over the past two decades, and significant progress has been
made in this subject \cite{4,5,6,7,8,9,10,11}. For example, two-dimensional
(2D) quantum spin Hall effects, also dubbed as 2D topological insulators,
have been observed in the CdTe/HgTe/CdTe sandwiched quantum well \cite{12}.


In two earlier independent studies, it has been demonstrated that the
spin-orbit coupling in the electronic band structures can lead to the
transverse spin current in the absence of impurity scattering \cite{13,14}.
This system is a metal and its spin Hall conductivity is a universal value
$q/8\pi$ that does not depend on physical details like the spin-orbit
coupling strength and electron mass when the Fermi surface is above the
two-fold degenerate point, where $q$ is the charge of an electron. Later works found that the dissipationless spin
current originates from interbranch transitions, and ladder diagrams can
cancel it by taking account into impurity weak scatterings \cite%
{15,16,17,18,19,20}. More explicitly, Rashba introduced a perpendicular
magnetic field to group the energies into Landau levels so that the
contribution of both intrabranch transitions and interbranch transitions to
spin Hall conductivity could be derived. It shows that the spin Hall
conductivity from the contribution of the above two transitions cancels out
as the magnetic field approaches zero. Therefore, the spin Hall conductivity vanishes due to the \textit{cancellation theorem} under zero magnetic
field \cite{21}.

Meanwhile, the rapid progress of ultracold atomic platforms has triggered
extensive studies of spin-orbit coupled quantum gases, including large-spin
systems \cite{22,23,24}. Recently, a universal intrinsic higher-rank
spin-tensor Hall (STH) effect has been proposed in pseudospin-1 ultracold
fermionic atoms beyond the scope of the conventional spin Hall effect in
spin-1/2 system \cite{25}. It is found that this effect could induce a
transverse spin tensor current driven by a longitudinal external electric
field, while no lower-rank spin current and charge current exist. In
particular, the higher-rank STH conductivity is a universal value $q/8\pi$ as
long as the Fermi energy level is located above the triply degenerate point.
However, this study only considers the contribution of interbranch
transitions and does not include the contribution of the intrabranch
transitions. Naturally, an interesting question arises: whether the
intrabranch transitions contribute to the higher-rank STH conductivity and
whether the cancellation theorem still holds for STH effect? More
critically, how do we expect to observe STH conductivity in a realistic
experiment setup?

In this paper, we address these crucial questions by studying 2D
pseudospin-1 quantum gases with intrinsic spin-tensor momentum coupling by
considering the contribution of both interbranch and intrabranch
transitions. We first introduce a perpendicular weak magnetic field into the
system, such that the continuous energies are grouped into Landau levels. We
analytically derive the conductivity from the contribution of both
intrabranch and interbranch transitions from discrete Landau levels and get
the conductivity as the magnetic field approaches 0. It is found that while
the spin Hall conductivity cancels out, the STH conductivity and the Hall
conductivity do not, where both contain a term that diverges inversely
proportional to the strength of the magnetic field, which resembles the Hall
conductivity in the integer quantum Hall effect. Taking a step further, we
discuss the effect of external Zeeman fields and observe resonance in
spin-tensor Hall conductivity due to the crossing of Landau levels.

The paper is organized as follows. In Sec. \ref{1}, we review the rank-2 STH
effect, which is the base model for our discussions. In Sec. \ref{2}, we
introduce a magnetic field into the model Hamiltonian, and present the
analytically derived Hall conductivity for charge, spin, and spin-tensor
degrees of freedom. In Sec. \ref{3}, we study the effect of the Zeeman
field, which can be tuned independently from the magnetic field in a
cold-atom system, and show that it leads to a resonance in the STH
conductivity. Finally, we conclude and discuss possible future directions in
Sec. \ref{4}.

\section{Review of higher-rank STH Effect}

\label{1} In this section, we give a brief review of higher-rank STH effect
with a focus on the special case of spin-1 and introduce relevant notations.

Following previous work \cite{25}, we consider a 2D \textquotedblleft
electron" gas consisting of spin-1 particles with charge $-q$ and an
effective mass $m^{\ast }$. Assuming the system is subject to a spin-tensor
momentum coupling (STMC), its physics is described by an effective
Halmitionian
\begin{equation}
H=\frac{\bm{\hat{p}}^{2}}{2m^{\ast }}-\frac{1}{\sqrt{2}}\frac{\alpha }{\hbar
}\left( \bm{\hat{\tau}}_{T}+\bm{\hat{\tau}}_{V}^{\ast }\right) \cdot \left( %
\bm{z}\times \bm{\hat{p}}\right) ,  \label{Eq1}
\end{equation}%
where $\alpha >0$ is the STMC strength, and $\bm{\hat{\tau}}_{T}=(\hat{%
\lambda}_{1},\hat{\lambda}_{2},\hat{\lambda}_{3})$ and $\bm{\hat{\tau}}_{V}=(%
\hat{\lambda}_{6},\hat{\lambda}_{7},\frac{\sqrt{3}}{2}\hat{\lambda}_{8}-%
\frac{1}{2}\hat{\lambda}_{3})$ are spin operators corresponding to SU(2)
subalgebras of SU(3). The operators $\hat{\lambda}_{i=1,...,8}$ take the
form of the Gell-Mann matrices $\lambda _{i=1,...,8}$ in natural spin basis $%
\left\{ \uparrow ,\nmid ,\downarrow \right\} $. See Appendix~\ref{notation}
for more details \cite{26}.

Diagonalizing the Hamiltonian in Eq.(\ref{Eq1}) yields three eigenstates,
which read (in basis $\left\{\uparrow,\nmid,\downarrow \right\}$)
\begin{equation*}
\left\vert \overline{1}k\right\rangle =%
\begin{pmatrix}
\frac{1}{2} \\
\frac{i}{\sqrt{2}}e^{i\theta_p} \\
\frac{1}{2}%
\end{pmatrix}%
, \left\vert 0k\right\rangle =%
\begin{pmatrix}
\frac{1}{\sqrt2} \\
0 \\
\frac{-1}{\sqrt2}%
\end{pmatrix}%
, \left\vert 1k\right\rangle =%
\begin{pmatrix}
\frac{1}{2} \\
\frac{-i}{\sqrt{2}}e^{i\theta_p} \\
\frac{1}{2}%
\end{pmatrix}%
,
\end{equation*}
with the corresponding eigenenergies $E_{\overline{1}k}=\frac{p^{2}}{2m^*}%
-\alpha k$, $E_{0k}=\frac{p^{2}}{2m^*}$ and $E_{1k}=\frac{p^{2}}{2m^*}%
+\alpha k$, where $\theta_p=\arctan(p_y/p_x)$ is the polar angle of the
momentum $\bm{p}$, $k$ is the magnitude of the wavevector, and $|\bm{p}|=\hbar k$. An example of
such a band structure is plotted in Fig. \ref{Fig1}(a) and the three energy
branches cross at zero momentum, leading to a triply degenerate point. In
the following, we assume that the Fermi energy $E_f$ is always higher than
the degenerate point.

In continuous space, the rank-2 STH conductivity can be computed using the
Kubo formula
\begin{equation}
\sigma_{xy}^{zz}=-q{\hbar}\int \frac {d^{2}k}{(2\pi)^{2}}{\Omega}_{xy}^{zz},
\label{sigmazz}
\end{equation}
with $\Omega_{xy}^{zz}=-\sum_{\lambda \neq \lambda^{^{\prime }}}(f_{\lambda
k}-f_{\lambda^{^{\prime }} k})\frac{\Im \left\langle \lambda k\vert \hat{J}%
_{2,x}^{zz}\vert\lambda^{^{\prime }} k\right\rangle\left\langle
\lambda^{^{\prime }} k\vert \hat{v}_{y}\vert\lambda k\right\rangle}{%
(E_{\lambda k}-E_{\lambda^{^{\prime }}k})^2}$ a generalized rank-2
spin-tensor Berry curvature. Here, $\lambda$ and $\lambda^{\prime }$ are
branch indices, $f_{\lambda k}=[e^{(E_{\lambda k}-E_{f})/k_{B}T}+1]^{-1}$ is
the Fermi-Dirac distribution function, $\bm{\hat{v}}=\partial_{\bm{\hat{p}}}%
\hat{H}$ is the velocity operator and $\hat{J}_{2,x}^{zz}=\frac{1}{2}%
\left\{P_{2},\hat{v}_{x}\right\}$ is the $x$ component of the rank-2 STH
current operator along $z$, where $P_{2}={\hbar}N_{zz}$ is the rank-2
spin-tensor polarization operator with diagonal $N_{zz}=\text{diag}(\frac{1}{%
3},\frac{-2}{3},\frac{1}{3})$ in natural basis.

After substituting the eigenstates and eigenenergies into the Kubo formula
in Eq.(\ref{sigmazz}), it is easily found that we have a non-zero rank-2 STH
conductivity from the transitions between the lowest branch and the highest
branch ($\lambda,\lambda^{\prime }=1,-1$)
\begin{eqnarray}
\sigma _{xy}^{zz}=\frac{q\hbar^{2}}{16\alpha m^{*}\pi}%
\int_{k_{F3}}^{k_{F1}}dk=\frac{q}{8\pi }
\end{eqnarray}
where $k_{F1,3}$ denotes the corresponding Fermi wavevectors. Note that the
relation $k_{F1}-k_{F3}=2\alpha m^* / \hbar^2$ has been applied in the above
calculations. Moreover, the STH conductivity is universal and does not depend
on any details of the physical system, such as the STMC strength and the
effective mass of the particles.

The rank-1 spin current operator can be defined similarly using the spin
polarization operator $F_z$. According to the Kubo formula, we can derive $%
\sigma_{xy}^{z}=0$. Taking a similar procedure, the rank-0 charge current
operator can be defined as $\hat{J}_0=-q\hat{v}_x$ and the corresponding
Hall conductivity also vanishes $\sigma_{xy}=0$. To summarize, this system
exhibits a universal STH conductivity, while both rank-1 spin Hall and rank-0
Hall conductivity vanishes. It produces the so-called STH effect, providing
an opportunity for generating dissipationless spin-tensor current and
potential device applications.

\section{Sum rules for higher-rank spin Hall conductivity}

\label{2} Similar to spin Hall effect in 2D electron gas with Ranshba
spin-orbit coupling \cite{14,15}, the original proposal of STH effect does
not consider the cancellation of the STH current in the presence of other
conditions like impurities or scattering \cite{25}. To gain a holistic view
of the problem, we follow the spirit of the previous study \cite{21} to
revisit the cancellation theorem in STH effect.

\subsection{Model Hamiltonian}

To start, we consider the STH effect model as in Eq.(\ref{Eq1}) subjected
to a perpendicular field $\bm{B}=B\bm{z}=\hat{\nabla}\times\bm{A}$ but
ignore the Zeeman term for now. The model Hamiltonian is then given by
\begin{equation}
\mathcal{\hat{H}} = \frac{\bm{\hat{\Pi}}^2}{2m^*}-\frac{1}{\sqrt{2}}\frac{%
\alpha}{\hbar}(\bm{\hat{\tau}}_T+\bm{\hat{\tau}}_V^*)\cdot(\bm{z}\times%
\mathbf{\hat{\Pi}}),
\end{equation}
where $\bm{\hat{\Pi}}=\bm{\hat{p}}+\frac{q}{c}\bm{A}$ is the kinetic
momentum.

\begin{figure}[t]
\centering
\includegraphics[width=0.45\textwidth]{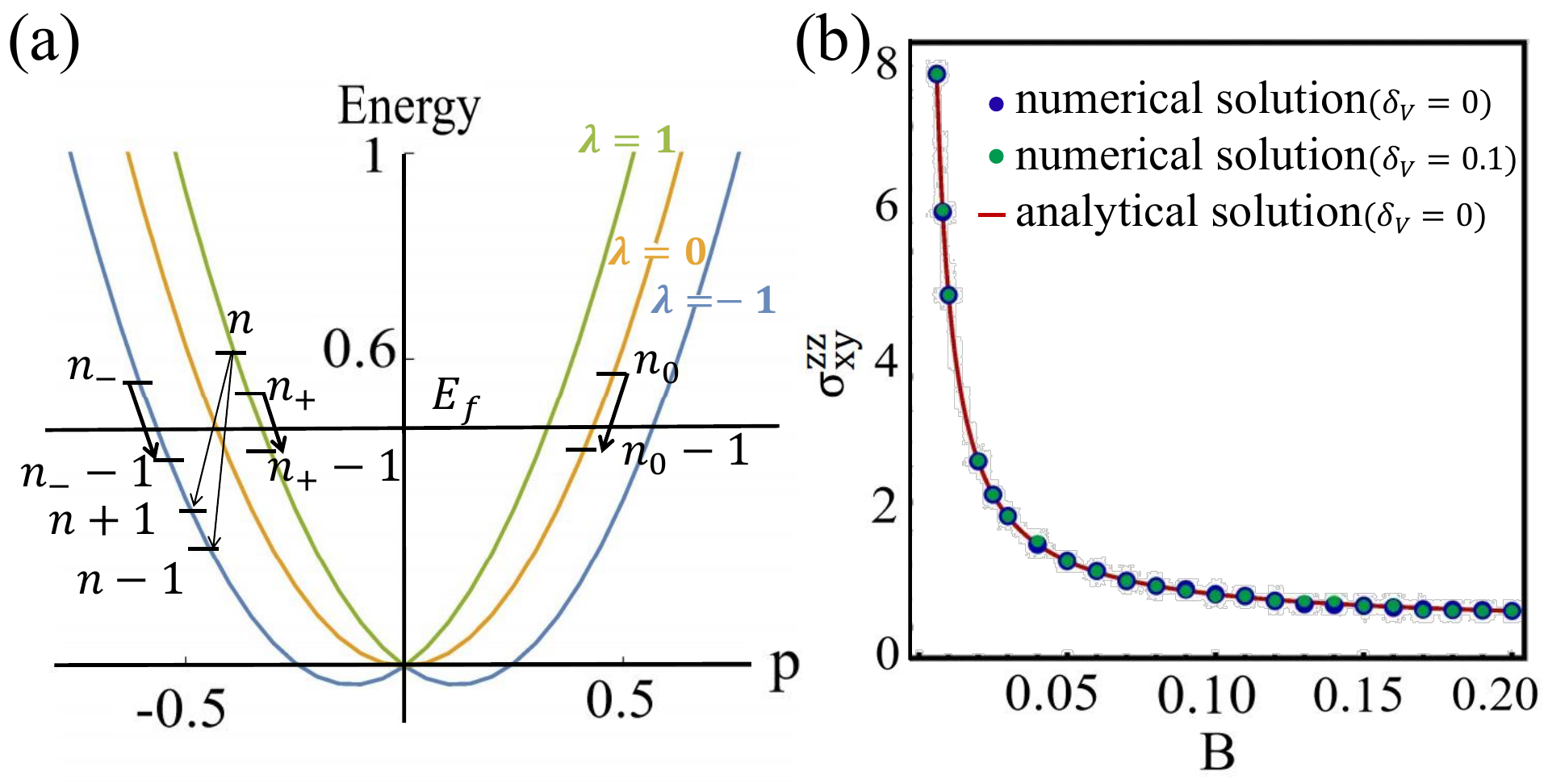} 
\caption{(a) Schematic illustration of the energy spectrum. Here three solid
lines with $\protect\lambda=\pm1, 0$ denote the energy spectrum. $n$ is the
quantum number of Landau levels for these branches. Different types of
transitions are labeled by arrows. Interbranch transition happens when $%
n\rightarrow n\pm1$ while the intrabranch transition corresponds to $n_{\protect%
\lambda}\rightarrow n_{\protect\lambda}-1$. The parameters are set to be $%
m^{\ast }=0.2, \protect\alpha =0.5$ and we use $\hbar$ as the units for
energy, $c$ for speed, and $q$ for charge throughout this paper. (b)
Numerical results with $\delta_{V}=0$ and $0.1$, and analytical results with $\delta_{V}=0$ for the rank-2 STH conductivity versus the
magnetic field. Common parameters used in the latter two panels are $m^{\ast
}=\protect\alpha=1$. }
\label{Fig1}
\end{figure}
Without loss of generality, we assume $B>0$ and define the bosonic operator $%
\hat{a}=\frac{1}{\sqrt{2}}\frac{l}{\hbar}(\hat{\Pi}_x-i\hat{\Pi}_y)$, where $%
l=\sqrt{\hbar c/|q|B}$ is the magnetic length. The Hamiltonian is then
transformed to 
$\mathcal{\hat{H}} = \hbar\omega_c\left(\hat{n}+\frac{1}{2}\right)+i\frac{%
\alpha}{l}\left[\hat{a}(\hat{V}_-+\hat{I}_+) -\hat{a}^\dagger(\hat{V}_++\hat{%
I}_-)\right]$, 
where $\hat{n}=\hat{a}^\dagger \hat{a}$ is the number operator, $%
\omega_c=qB/m^*c$ is cyclotron frequency and $\hat{I}_\pm=(\hat{\tau}%
_{T,x}\pm i\hat{\tau}_{T,y})/2$ and $\hat{V}_\pm=(\hat{\tau}_{V,x}\pm i\hat{%
\tau}_{V,y})/2$ are ladder generators of the SU(3) group.

We notice that the mode $\left|\nmid,n=0\right\rangle$ is an eigenstate
(Here $\left\{\uparrow,\nmid,\downarrow \right\}$ represent the three spin
states)
\begin{equation}
\mathcal{\hat{H}}\left|\nmid,0\right\rangle=\frac{1}{2}\hbar\omega_c\left|%
\nmid,0\right\rangle,
\end{equation}
which is not coupled to any other states.

For a given $n>0$, three modes $\left\vert \uparrow ,n-1\right\rangle $, $%
\left\vert \downarrow ,n-1\right\rangle $ and $\left\vert \nmid
,n\right\rangle $ couple with each other, but decouple from other modes \cite%
{27}. The subspace $\left\{ \left\vert \uparrow ,n-1\right\rangle
,\left\vert \nmid ,n\right\rangle ,\left\vert \downarrow ,n-1\right\rangle
\right\} $ form an effective spin-1 system $\{|\tilde{\uparrow}\rangle ,|%
\tilde{\nmid}\rangle ,|\tilde{\downarrow}\rangle \}$, within which the
Hamiltonian $\mathcal{\hat{H}}$ can be expressed as
\begin{equation}
\mathcal{H}_{Z}=n\hbar \omega _{c}+\frac{1}{2}(1-2\tilde{F}_{z}^{2})\hbar
\omega _{c}-\sqrt{n}\frac{\alpha }{l}\left( \tilde{\tau}_{T,y}+\tilde{\tau}%
_{V,y}^{\ast }\right) ,
\end{equation}%
where $\tilde{F}_{z}$, $\tilde{\tau}_{T,y}$ and $\tilde{\tau}_{V,y}$ represent the unfolding form of $F_{z}$, $\tau _{T,y}$ and $\tau _{V,y}$ under the space of spin and particle numbers. They take the same matrix forms. We have
three eigenvalues labeled by $\lambda =\pm 1,0$ as
\begin{eqnarray}
E_{\lambda n} &=&\hbar \omega _{c}(n+\lambda c_{n}), \\
E_{0n} &=&\hbar \omega _{c}(n-\frac{1}{2}),
\end{eqnarray}%
where $c_{n}=\sqrt{\gamma ^{2}n+\frac{1}{4}}$ and $\gamma =[2(m^{\ast
}\alpha ^{2})/\hbar ^{3}\omega _{c}]^{1/2}$ is a dimensionless constant
representing the relative strength of STMC and cyclotron frequency. Some
examples of the corresponding Landau levels are labeled in Fig. \ref{Fig1}%
(a). The corresponding wave functions read (in basis $\{|\tilde{\uparrow}%
\rangle ,|\tilde{\nmid}\rangle ,|\tilde{\downarrow}\rangle \}$)
\begin{eqnarray}
\left\vert \lambda n\right\rangle  &=&(\frac{1}{\sqrt{2}}b_{n}^{\bar{\lambda}%
},i\lambda b_{n}^{\lambda },\frac{1}{\sqrt{2}}b_{n}^{\bar{\lambda}})^{T},
\label{Eq8} \\
\left\vert 0n\right\rangle  &=&\frac{1}{\sqrt{2}}(-1,0,1)^{T}.
\end{eqnarray}%
In Eq.(\ref{Eq8}), $b_{n}^{\lambda }=\frac{1}{\sqrt{2}}\sqrt{1+\frac{%
\lambda }{2c_{n}}}$, and $\overline{\lambda }=-\lambda $ with $\lambda =\pm 1
$. Obviously, $|0n\rangle $ is a dark state decoupled from the other states
and remains as an even superposition of the two spin components $\left\vert
\uparrow \right\rangle $ and $\left\vert \downarrow \right\rangle $ at
different Landau levels $n$.

\subsection{Rank-2 STH Conductivity}

Similarly, we use the standard Kubo-Greenwood formula to compute the rank-2
STH Hall conductivity
\begin{equation}
\sigma_{xy}^{zz}(\omega)=-i\frac{q}{\pi\hbar l^2}\sum_{\lambda
n\lambda^{\prime }n^{\prime }}\xi\frac{\langle \lambda n\vert \hat{v}_y\vert
\lambda^{\prime }n^{\prime }\rangle\langle \lambda^{\prime }n^{\prime }\vert
\hat{J}^{zz}_{2,x}\vert \lambda n\rangle}{(\omega_{\lambda
n}-\omega_{\lambda^{\prime }n^{\prime }})^2-\omega^{2}},  \label{eq8}
\end{equation}
where $\xi=f_{\lambda n} - f_{\lambda^{\prime }n^{\prime }}$, $f_{\lambda
n}=1/(e^{(E_{\lambda n}-E_{f}/k_{B}T}$+1) is the Fermi-Dirac distribution function.
The sum requires $\hbar\omega_{\lambda n}>E_f, \hbar\omega_{\lambda^{\prime
}n^{\prime }}<E_f$, where $E_f$ is the Fermi energy. In this work, we
consider the static $\omega=0$ limit with the temperature $T=0$. We assume
that the Fermi energy is large enough $E_f = \hbar\omega_f \gg\hbar\omega_c$%
. Then the Fermi momentum becomes $k_{\pm}^2l^2=2\eta+{\gamma}^{2}\mp\sqrt{%
\gamma^{4}+4\eta\gamma^2+1}$ and $k_{0}^2l^2=2\eta+1$, where $%
\eta=\omega_{f}/\omega_{c} \gg 1$.

The velocity operator is defined as $\hat{v}_y=\frac{i}{\hbar}[\mathcal{\hat{%
H}},\hat{y}]$, where $\hat{y}=-l^2\hat{k}_x$ and the guiding center operator
is neglected. The momentum operator $\mathbf{\hat{\Pi}}=\hbar \bm{\hat{k}}$
can be related to creation and annihilation operators by $\hat{k}_x=\frac{(%
\hat{a}^\dagger+\hat{a})}{\sqrt{2}l}$ and $\hat{k}_y=\frac{(\hat{a}^\dagger-%
\hat{a})}{\sqrt{2}li}$. Then, we have
\begin{equation}
\langle \lambda n\vert \hat{v}_y\vert \lambda^{\prime }n^{\prime
}\rangle=(\omega_{\lambda n}-\omega_{\lambda^{\prime }n^{\prime }})\langle \lambda
n\vert \hat{k}_x\vert \lambda^{\prime }n^{\prime }\rangle.
\end{equation}

The higher-rank spin-tensor current operator is defined similarly to that in
the previous section and it can be written as
\begin{equation}
\hat{J}_{2,x}^{zz}=\frac{\hbar ^{2}\hat{k}_{x}}{m}\hat{N}_{zz}+\frac{\alpha
}{6\sqrt{2}}(\hat{\lambda}_{2}+\hat{\lambda}_{7}^{\ast }).
\end{equation}%
Unlike the case of Rashba spin-orbit coupling, one can prove that it is
impossible to find a Hermitian operator $\hat{Q}$ satisfying $[\mathcal{\hat{%
H}},\hat{Q}]=\hat{J}_{2,x}^{zz}$ due to the tensor component. Thus, we need
to evaluate each term individually.

Even under the non-zero magnetic field, we can still show that only the
transitions between the highest band ($\lambda=1$) and the lowest band ($%
\lambda=-1$) have non-zero contributions to the rank-2 STH conductivity if
only interbranch transitions are considered. To derive the analytical
results for STH conductivity from the contribution of interbranch
transitions, we write down the Taylor expansion of the Kubo formula in terms
of $\eta$ and keep only the leading order. The rank-2 STH conductivity then
becomes $\sigma_{xy, inter}^{zz} = q/8\pi$, which is consistent with the
results obtained from the continuous limit in Sec.~\ref{1} (See Appendix \ref%
{notation1} for more details). We also calculate the rank-2 STH conductivity from the contribution of interbranch transitions by treating the magnetic field as a perturbation. The result is consistent with the above obtained result (See Appendix \ref%
{notation4} for more details).

Next, we consider the correction from intrabranch transitions to $%
\sigma_{xy}^{zz}$. The practice is similar to that of interbranch
transitions. We find that only the intrabranch transitions between energy
levels near the Fermi surface have meaningful contributions.
We first focus on the case $\lambda=1$ and $\lambda^{\prime }=-1$, and have
the rank-2 STH conductivity from the contribution of intrabranch $\lambda=1$
and $\lambda^{\prime }=-1$ (see Appendix~\ref{notation1} for more details)
\begin{eqnarray}
\sigma_{xy,intra,\pm 1}^{zz}&=&-\frac{q}{\pi\hbar}\frac{\langle \lambda
n_{\lambda}\vert \hat{k}_{x}\vert \lambda n_{\lambda}-1\rangle\langle
\lambda n_{\lambda}-1\vert \hat{J}_{2,x}^{zz} \vert \lambda
n_{\lambda}\rangle}{\omega_{n_{\lambda}}-\omega_{n_{\lambda}-1}}  \notag \\
&=&\frac{q\eta}{6\pi}+\frac{q(-5+2\gamma^2)}{24\pi}+\frac{%
q(3-4\gamma^2+\gamma^4) }{96\pi \gamma^2 \eta}.
\end{eqnarray}
Besides, we also have a non-vanishing contribution $\sigma_{xy,intra,0}^{zz}
=-\frac{q\hbar (l^2 k_{0}^2 -2)}{12m^*\pi l^{2}\omega_c}$ from the
intrabranch transitions within $\lambda=0$. So the total intrabranch STH
conductivity reads $\sigma_{xy,intra}^{zz}=-\frac{q}{8\pi}+\frac{q}{12\pi}%
\gamma^{2}$. Lastly, we arrive at the final rank-2 STH {conductivity by
summing both interbranch and intrabranch contributions
\begin{eqnarray}
\sigma _{xy}^{zz} &=&\sigma _{xy,inter}^{zz}+\sigma _{xy,intra}^{zz} =\frac{%
m^{\ast 2}\alpha ^{2}c}{6\pi \hbar ^{3}B},
\end{eqnarray}
which is plotted as the solid curve in Fig. \ref{Fig1}(b). In above calculations, higher-order small quantities on $B$ have been neglected under $B\ll 1$ and $\eta \gg 1$. In the following we further use numerical methods to evaluate $\sigma _{xy}^{zz}$ using Eq.(\ref{eq8}) exactly, and the results are plotted by blue dots as in Fig. \ref{Fig1}(b). We observe the numerical solution is in excellent agreement with the analytical result when $B\ll 1$.

To understand the origin of such a divergence of STH conductivity we project $%
\mathcal{\hat{H}}$ onto a pseudospin subspace $\{\left\vert \Uparrow
\right\rangle ,\left\vert \Downarrow \right\rangle \}$, where the
pseudospins $\left\vert \Uparrow \right\rangle =\frac{1}{\sqrt{2}}%
(\left\vert \tilde{\uparrow}\right\rangle +\left\vert \tilde{\downarrow}%
\right\rangle )$ and $\left\vert \Downarrow \right\rangle =\left\vert \tilde{%
\nmid}\right\rangle $ are orthogonal to $\left\vert \o \right\rangle =\frac{1%
}{\sqrt{2}}(\left\vert \tilde{\uparrow}\right\rangle -\left\vert \tilde{%
\downarrow}\right\rangle )$. On the new basis, the effective Hamiltonian
shows a Rashba spin-orbit coupling and there is a constant term emerging in
the effective current operator, which relates the divergence in STH
conductivity to that of the Hall conductivity (see Appendix~\ref{notation3}
for more details).




We have also carefully examined the rank-1 spin Hall and rank-0 Hall
conductivity. Since the contribution to rank-1 spin Hall conductivity from
both intrabranch and interbranch transitions is zero, spin Hall conductivity
vanishes. While the contribution to rank-0 Hall conductivity from the
interbranch transitions is zero, the non-zero contribution from the
intrabranch transitions gives the Hall conductivity $\sigma_{xy}\sim1/B$,
showing divergent behavior as $B$ approaches zero (see Appendix~\ref%
{notation1} for more details). We also calculate the Hall conductivity in
spin-1/2 system. The result is similar to that in spin-1 system, where only
intrabranch transitions contribute to the Hall conductivity inversely
proportional to $1/B$ (see Appendix~\ref{notation2} for more details).


\section{Resonance in STH conductivity under the Zeeman field}

\label{3}
\begin{figure}[t]
\par
\includegraphics[width=0.5\textwidth]{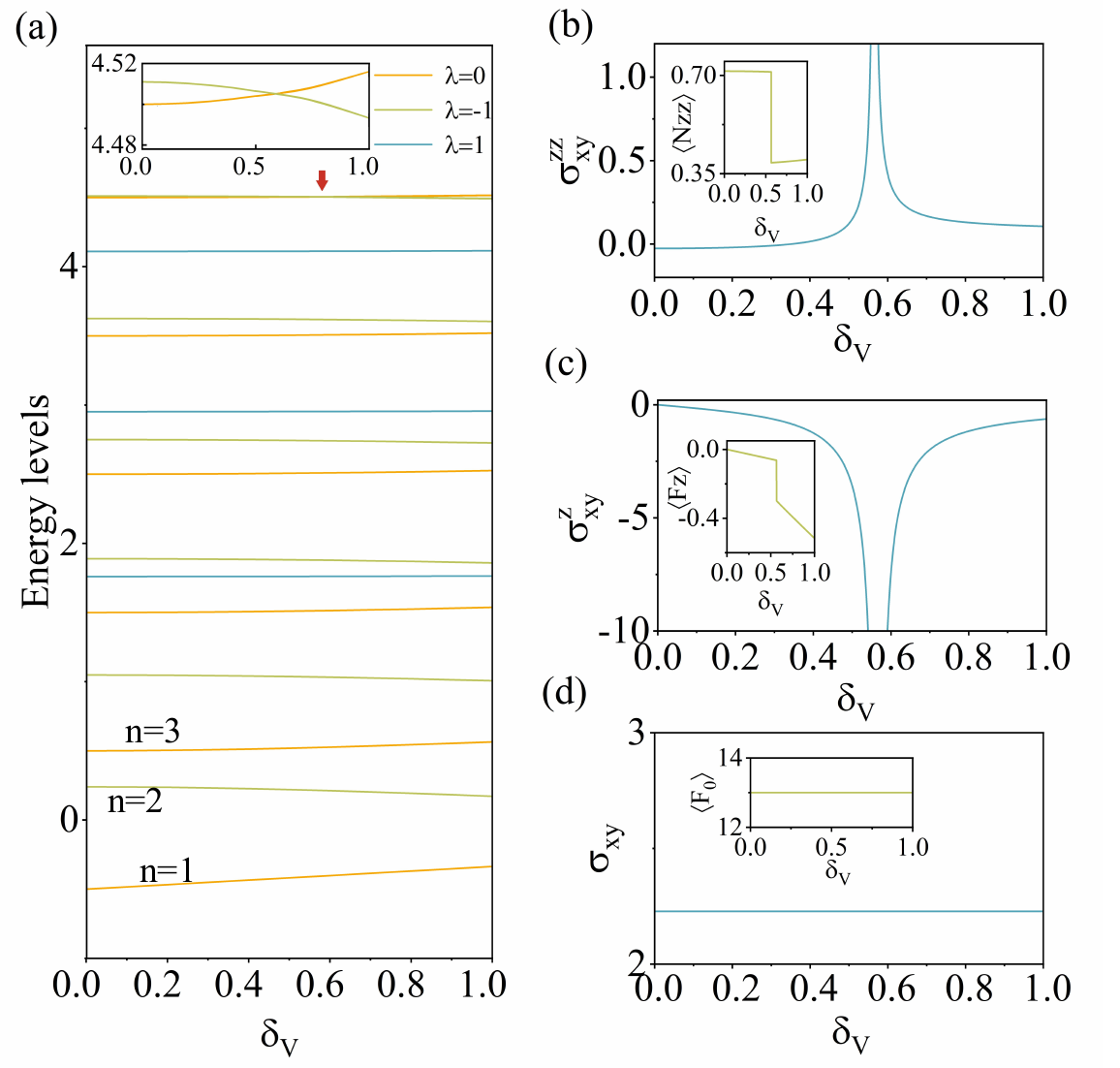}
\caption{Numerical results for rank-2 spin-tensor, rank-1 spin and rank-0
charge Hall conductivity under the Zeeman field. (a) Landau levels of an
electron as functions of $\protect\delta_V$. The inset zooms in the Landau
level crossing indicated by the red arrow in main figure. (b) Numerical
results for rank-2 STH conductivity under a spin-vector Zeeman field $\protect%
\delta_V$. [(c)-(d)] Numerical results for rank-1 spin and rank-0 Hall
conductivity under a spin-vector Zeeman field $\protect\delta_V$,
respectively. In the plots, we fix Fermi energy $E_{f}$ at the energy
crossing, $m^{\ast }=\protect\alpha=1$, and $B=6.1$ 
}
\label{Fig2}
\end{figure}

The intrabranch transitions lead to divergent rank-2 spin-tensor and rank-0
charge Hall conductivity when $B$ approaches zero in the absence of Zeeman field. In the following, we would
like to consider the system under the Zeeman field, i.e., with a finite spin-vector term $%
\delta _{V}\hat{F}_{z}$, where $\delta _{V}\neq0$. In this case, the triply degenerate point would be lifted \cite{25}. However, the rank-2 STH conductivity also exhibits divergent behavior as $B$ approaches zero if the Fermi energy is above the triply degenerate point, as demonstrated by numerical results in Fig. \ref{Fig1}(b), similar to the case with $\delta _{V}=0$. We note that since the energy levels are split between Landau
levels under finite magnetic field, the divergent spin-tensor conductivity from the contribution of
intrabranch transitions vanishes. Furthermore, due to the introduction of the
spin-vector term, the Landau levels from different energy branches could
cross as $\delta _{V}$ increases, leading to that the STH conductivity may
exhibit interesting resonance behavior. In the following, we examine such a
resonance in STH conductivity. For simplicity, we assume that the Zeeman
fields could be tuned independently of the external magnetic field, which
can be realized via cold atoms as in the original proposal \cite{25}.


In the presence of Zeeman fields, the triply-degenerate point shown in Fig. %
\ref{Fig1}(a) is broken. It becomes impractical to derive analytic results
in general. So we numerically calculate the energies of the system and STH
conductivity by the Kubo formula. The Landau levels of the system with $%
\delta _{V}\hat{F}_{z}$ have been shown in Fig. \ref{Fig2}(a) in terms of
energy branch. There is always an energy split between adjacent Landau
levels from the same energy branch. As $\delta _{V}$ increases, two Landau
levels in $\lambda =0$ and $\lambda =-1$ branches cross at the critical
value ${\delta _{V}}_{c}=0.56$, and forms a degenerate point as in the inset
of Fig. \ref{Fig2}(a). We also compute the STH conductivity versus $\delta
_{V}$ with the Fermi energy fixed at the degenerate point, as in Fig. \ref%
{Fig2}(b). It is obvious that a resonance of STH conductivity occurs at the
critical point. This can be understood from two aspects. Firstly, for ${%
\delta _{V}}<{\delta _{V}}_{c}$, the Landau level in branch $\lambda =0$ is
below that in branch $\lambda =-1$. While for ${\delta _{V}}>{\delta _{V}}%
_{c}$, these two Landau levels reverse. Since the states in these two
branches carry different values of $\langle {\hat{N}_{zz}}\rangle $, the
total $\langle {\hat{N}_{zz}}\rangle $ has a jump at the Landau level
crossing point, as shown in the inset of Fig. \ref{Fig2}(b). Secondly, the
denominator of the Kubo formula as in Eq. (\ref{eq8}) becomes zero at the
critical point, and $\sigma _{xy}^{zz}$ would be divergent. Therefore, there
is a resonance of STH conductivity at the crossing point.

In summary, STH conductivity exhibits divergence in the absence of the Zeeman field. However, in the presence of Zeeman field, STH conductivity becomes finite in general and exhibits divergent resonance when the Landau levels cross at the Fermi energy. We have also examined rank-1 and
rank-0 Hall conductivity. $\langle {\hat{F}_{z}}\rangle $ for the system with
spin vector term have a jump, and the corresponding rank-1 Hall conductivity
becomes also divergent, leading to a resonance as shown in Fig. \ref{Fig2}%
(c). While as in Fig. \ref{Fig2}(d), $\langle {\hat{F}_{z}}\rangle $has no
jump, and there is no resonance in rank-0 Hall conductivity.


\section{Conclusion and outlook}

\label{4} In this work, we have revisited the cancellation theorem in
higher-rank STH effects. We take the minimal toy model with a rank-2 STH
effect in continuous space and introduce an external magnetic field into the
model to study the cancellation of interbranch contribution from intrabranch
contribution. For the toy model in continuous space, we can only compute the
interbranch contributions and only the rank-2 STH conductivity is quantized
to $q/8\pi$ while both the charge and spin conductances vanish. When
intrabranch transitions are included, for the rank-2 STH Hall conductivity
the intrabranch contribution cancels the interbranch contribution but brings
up an extra term that is also proportional to $1/B$, similar to the ordinary
quantum Hall effect with $B$ approaching zero: In the quantum Hall effect, a
series of quantized resistance platforms $R_H=h/iq^2$ ($i$ is an integer)
appear as the Hall resistance changes with the magnetic field. If the
intrabranch transitions are included, the rank-1 spin Hall conductivity
always vanish but the rank-0 charge Hall conductivity becomes $\sigma_{xy}=%
q^{2}(-3+2\gamma^{2}+6\eta)/4\pi\hbar$ that is proportional to $1/B$%
. This is due to the transverse charge current generated by the magnetic
field, similar to the Hall conductivity in the Hall effect \cite%
{30,31,32}. The system in this work could be realized in a ultracold atomic system with good controllability \cite{25,33,34,35,36,37,38}. We can measure the rank-2 STH conductivity via spin accumulation in ultracold-atom experiments \cite{25}. The measurement of STH conductivity resonance only needs to observe if $\langle {\hat{N}_{zz}}\rangle $ has a jump as the spin-vector Zeeman field $\protect\delta_V$ increases.

When the magnetic field becomes weak, the filling number of Landau level increases and the Hall conductivity increases continuously \cite{39,40,41}.
As the magnetic field approaches zero, the particle's cyclotron radius is infinite similar to the case of the usual quantum Hall effect magnetic field, and the opposite boundary is reached, resulting in the spin-tensor Hall effect not being observed in a finite size sample. We also
have studied the system under the Zeeman field. Interestingly, there is an
observable resonance in rank-2 STH conductivity since the landau level
crossing. In this study the potential disorders
are not considered. They could broaden Landau levels and lead to localization of "electrons". We expect the energy gap at Landau level
crossing is negligible in the presence of weak disorders, so that the resonance of STH conductivity could also be observed. The detailed analysis is
beyond the scope of this study and would be left for the further study.

\begin{acknowledgments}
We thank C. Zhang and E. I. Rashba for inspiring discussion. X. He and Y. Wu
are supported by NSFC under the grant No.12275203, Innovation Capability
Support Program of Shaanxi (2022KJXX-42), 2022 Shaanxi University Youth
Innovation Team Project (K20220186).
\end{acknowledgments}

\appendix

\section{Definition of spin operators in a spin-1 system}

\label{notation} Throughout this work, we assume the natural basis for
different spins
\begin{equation}
\left\vert \uparrow \right\rangle =%
\begin{pmatrix}
1 \\
0 \\
0%
\end{pmatrix}%
,\left\vert \nmid \right\rangle =%
\begin{pmatrix}
0 \\
1 \\
0%
\end{pmatrix}%
,\left\vert \downarrow \right\rangle =%
\begin{pmatrix}
0 \\
0 \\
1%
\end{pmatrix}%
.
\end{equation}%
The matrix forms for operators of spin vectors in the above basis are given
by
\begin{equation}
F_{x}=\frac{1}{\sqrt{2}}%
\begin{pmatrix}
0 & 1 & 0 \\
1 & 0 & 1 \\
0 & 1 & 0%
\end{pmatrix}%
,F_{y}=\frac{1}{\sqrt{2}}%
\begin{pmatrix}
0 & -i & 0 \\
i & 0 & -i \\
0 & i & 0%
\end{pmatrix}%
\end{equation}%
and a diagonal matrix $F_{z}=\text{diag}(1,0,-1)$. The Gell-Mann matrices
are defined as
\begin{eqnarray}
\lambda _{1} &=&%
\begin{pmatrix}
0 & 1 & 0 \\
1 & 0 & 0 \\
0 & 0 & 0%
\end{pmatrix}%
,\lambda _{2}=%
\begin{pmatrix}
0 & -i & 0 \\
i & 0 & 0 \\
0 & 0 & 0%
\end{pmatrix}%
,\lambda _{3}=%
\begin{pmatrix}
1 & 0 & 0 \\
0 & -1 & 0 \\
0 & 0 & 0%
\end{pmatrix}%
,  \notag \\
\lambda _{4} &=&%
\begin{pmatrix}
0 & 0 & 1 \\
0 & 0 & 0 \\
1 & 0 & 0%
\end{pmatrix}%
,\lambda _{5}=%
\begin{pmatrix}
0 & 0 & -i \\
0 & 0 & 0 \\
i & 0 & 0%
\end{pmatrix}%
,\lambda _{6}=%
\begin{pmatrix}
0 & 0 & 0 \\
0 & 0 & 1 \\
0 & 1 & 0%
\end{pmatrix}%
,  \notag \\
\lambda _{7} &=&%
\begin{pmatrix}
0 & 0 & 0 \\
0 & 0 & -i \\
0 & i & 0%
\end{pmatrix}%
,\lambda _{8}=\frac{1}{\sqrt{3}}%
\begin{pmatrix}
1 & 0 & 0 \\
0 & 1 & 0 \\
0 & 0 & -2%
\end{pmatrix}%
.
\end{eqnarray}




\section{Different-rank Hall conductivity in a spin-1 system}

\label{notation1}

\subsection{Rank-2 STH conductivity}

In the following, we will provide more details on the analytical derivation
of the higher-ranker STH conductivity discussed in the main text. The rank-2 STH conductivity consists of two parts from the contribution of interbranch- and intrabranch transitions.

First of all,
we calculate the conductivity from the contribution of interbranch
transitions as follows,
\begin{eqnarray}
\sigma _{xy,inter}^{zz} &=&-\frac{q}{\pi \hbar }\sum_{\lambda n\lambda
^{\prime }n^{\prime }}\frac{\langle \lambda n|\hat{k}_{x}|\lambda ^{\prime
}n^{\prime }\rangle \langle \lambda ^{\prime }n^{\prime }|\hat{J}%
_{2,x}^{zz}|\lambda n\rangle }{\omega _{\lambda n}-\omega _{\lambda ^{\prime
}n^{\prime }}} \notag \\
&=&d\sum_{n}\left( \frac{A_{n}}{\omega _{1n}-\omega _{\overline{1}n+1}}+%
\frac{B_{n}}{\omega _{1n}-\omega _{\overline{1}n-1}}\right) .  \label{BB5}
\end{eqnarray}%
Here, $d=-\frac{q\hbar }{2m^{\ast }\pi l^{2}}$, $A_{n}=E_{\hat{a}}E_{\hat{a}%
^{\dagger }\hat{N}_{zz}}^{\prime }+\frac{m^{\ast }l\alpha }{3h^{2}}E_{\hat{a}%
}E_{\hat{\lambda}_{2}}^{\prime }$, $B_{n}=E_{\hat{a}^{\dagger }}E_{\hat{a}%
\hat{N}_{zz}}^{\prime }+\frac{2m^{\ast }l\alpha }{3h^{2}}E_{\hat{a}^{\dagger
}}E_{\hat{\lambda}_{2}}^{\prime }$, where $E_{\hat{O}}=\langle \lambda n|%
\hat{O}|\lambda ^{\prime }n^{\prime }\rangle $, $E_{\hat{O}}^{\prime
}=\langle \lambda ^{\prime }n^{\prime }|\hat{O}|\lambda n\rangle $ with the
operator $\hat{O}=\hat{a}$, $\hat{a}^{\dagger }\hat{N}_{zz}$, $\hat{\lambda}%
_{2}$, $\hat{a}^{\dagger }$, $\hat{a}\hat{N}_{zz}$, and

\begin{eqnarray*}
E_{\hat{a}}E_{\hat{a}^{\dagger }\hat{N}_{zz}}^{\prime } &=&\frac{1}{%
48c_{n+1}c_{n}}\left[ n(2c_{n+1}+1)(2c_{n}-1)-2\times \right.  \\
&&\left. (n-1)(2c_{n+1}-1)(2c_{n}+1)+4n(n+1)\gamma ^{2}\right] , \\
E_{\hat{a}^{\dagger }}E_{\hat{a}\hat{N}_{zz}}^{\prime } &=&\frac{1}{%
48c_{n-1}c_{n}}[(n-1)(2c_{n-1}+1)(2c_{n}-1)- \\
&&\left. 2n(2c_{n-1}-1)(2c_{n}+1)+4n(n-1)\gamma ^{2}\right] , \\
E_{\hat{a}}E_{\hat{\lambda}_{2}}^{\prime } &=&\frac{1}{4\sqrt{2}}[n\frac{%
\gamma (2c_{n+1}+1)}{2c_{n+1}c_{n}}-(n+1)\frac{\gamma (2c_{n}+1)}{%
2c_{n+1}c_{n}}], \\
E_{\hat{a}^{\dagger }}E_{\hat{\lambda}_{2}}^{\prime } &=&\frac{1}{4\sqrt{2}}%
[(n-1)\frac{\gamma (2c_{n}-1)}{2c_{n-1}c_{n}}-n\frac{\gamma (2c_{n-1}-1)}{%
2c_{n-1}c_{n}}].
\end{eqnarray*}%
In the above calculation, we have used following relations. When $\hat{O}=%
\hat{a}$ or $\hat{O}=\hat{a}^{\dagger }$, the non-zero terms are
\begin{equation}
\langle 1n|\hat{O}|\overline{1}n\pm 1\rangle =\pm \sqrt{n}b_{n\pm 1}^{\pm
}b_{n}^{\mp }\mp \sqrt{n\pm 1}b_{n}^{\pm }b_{n\pm 1}^{\mp }.
\end{equation}%
When $\hat{O}=\hat{\lambda}_{2}$, we have
\begin{equation}
\langle \overline{1}n\pm 1|\hat{O}|1n\rangle =\mp \frac{1}{\sqrt{2}}b_{n\pm
1}^{\pm }b_{n}^{\pm }.
\end{equation}%
The case with $\hat{O}=\hat{a}\hat{N}_{zz}$ is more complicated and the
non-zero terms read
\begin{equation}
\langle \overline{1}n-1|\hat{a}\hat{N}_{zz}|1n\rangle =\frac{1}{3}[\sqrt{n-1}%
b_{n}^{-}b_{n-1}^{+}+2\sqrt{n}b_{n}^{+}b_{n-1}^{-}].
\end{equation}%
Similarly, for $\hat{O}=\hat{a}^{\dagger }\hat{N}_{zz}$, we have
\begin{equation}
\langle \overline{1}n+1|\hat{a}^{\dagger }\hat{N}_{zz}|1n\rangle =\frac{1}{3}%
[\sqrt{n}b_{n}^{-}b_{n+1}^{+}+2\sqrt{n+1}b_{n}^{+}b_{n+1}^{-}].
\end{equation}%
After substituting the detailed results of $A_{n}$ and $B_{n}$ into Eq.(\ref%
{BB5}) and performing a Taylor expansion on $\eta $, we arrive at a simple
equation,
\begin{eqnarray}
\sigma _{xy,inter}^{zz} &=&-\frac{q}{\pi \hbar }\sum_{\lambda n\lambda
^{\prime }n^{\prime }}\frac{\langle \lambda n|\hat{k}_{x}|\lambda ^{\prime
}n^{\prime }\rangle \langle \lambda ^{\prime }n^{\prime }|\hat{J}%
_{2,x}^{zz}|\lambda n\rangle }{\omega _{\lambda n}-\omega _{\lambda ^{\prime
}n^{\prime }}}  \notag \\
&=&\frac{q}{8\pi }.
\end{eqnarray}

Next we derive rank-2 STH conductivity from the contribution of intrabranch
transitions. This part consists of two terms. At first, we calculate the
contribution from $\lambda =\pm 1$ intrabranch transitions. Conclusions are
as follows, 
\begin{eqnarray}
\sigma _{xy,intra,\pm 1}^{zz} &=&-\frac{q}{\pi \hbar }\frac{\langle \lambda
n_{\lambda }|\hat{k}_{x}|\lambda n_{\lambda }-1\rangle \langle \lambda
n_{\lambda }-1|\hat{J}_{2,x}^{zz}|\lambda n_{\lambda }\rangle }{\omega
_{n_{\lambda }}-\omega _{n_{\lambda }-1}}  \notag  \label{BB6} \\
&=&d\frac{T_{\hat{a}^{\dagger }}T_{\hat{a}\hat{N}_{zz}}^{\prime }}{\omega
_{n_{\lambda }}-\omega _{n_{\lambda }-1}}-\frac{q\alpha }{6\pi \hbar l}\frac{%
T_{\hat{a}^{\dagger }}T_{\hat{\lambda}_{2}}^{\prime }}{\omega _{n_{\lambda
}}-\omega _{n_{\lambda }-1}},
\end{eqnarray}%
where
\begin{eqnarray}
\frac{T_{\hat{a}^{\dagger }}T_{\hat{a}\hat{N}_{zz}}^{\prime }}{\omega
_{n_{\lambda }}-\omega _{n_{\lambda }-1}} &=&\lambda \frac{(-1+\gamma ^{2})%
\sqrt{\eta }}{4\gamma \omega _{c}}+\frac{5-4\gamma ^{2}-4\eta }{24\omega _{c}%
}+  \notag \\
&&\lambda \frac{(3+3\gamma ^{2}-5\gamma ^{4}+5\gamma ^{6})\sqrt{\frac{1}{%
\eta }}}{96\gamma ^{3}\omega _{c}}-  \notag \\
&&\frac{(3-4\gamma ^{2}+\gamma ^{4})}{96\gamma ^{2}\omega _{c}\eta }+O({%
\frac{1}{\eta }})^{\frac{3}{2}}, \\
\frac{T_{\hat{a}^{\dagger }}T_{\hat{\lambda}_{2}}^{\prime }}{\omega
_{n_{\lambda }}-\omega _{n_{\lambda }-1}} &=&\overline{\lambda }\frac{\sqrt{%
\eta }}{2\sqrt{2}\omega _{c}}+\frac{\gamma }{2\sqrt{2}\omega _{c}}+\frac{%
\lambda }{16\sqrt{2\eta }\gamma ^{2}\omega _{c}}\times   \notag \\
&&(1+2\gamma ^{2}+3\gamma ^{4})+O({\frac{1}{\eta }})
\end{eqnarray}%
with $T_{\hat{O}}=\langle \lambda n_{\lambda }|\hat{O}|\lambda n_{\lambda
}-1\rangle $, $T_{\hat{O}}^{\prime }=\langle \lambda n_{\lambda }-1|\hat{O}%
|\lambda n_{\lambda }\rangle $, and $\hat{O}=\hat{a}^{\dagger }$, $\hat{a}%
\hat{N}_{zz}$, $\hat{\lambda}_{2}$. 
After substituting above equations into Eq.(\ref{BB6}) and performing a
Taylor expansion on $\eta $, we get the STH conductivity from the
contribution of intrabranch transitions for $\lambda =\pm 1$,
\begin{eqnarray}
\sigma _{xy,intra,\pm 1}^{zz} &=&-\frac{q}{\pi \hbar }\frac{\langle \lambda
n_{\lambda }|\hat{k}_{x}|\lambda n_{\lambda }-1\rangle \langle \lambda
n_{\lambda }-1|\hat{J}_{2,x}^{zz}|\lambda n_{\lambda }\rangle }{\omega
_{n_{\lambda }}-\omega _{n_{\lambda }-1}}  \notag \\
&=&\frac{q\eta }{6\pi }+\frac{q(-5+4\gamma ^{2})}{24\pi }+\frac{q(3-4\gamma
^{2}+\gamma ^{4})}{96\pi \gamma ^{2}\eta }
\end{eqnarray}%
Similarly, the contribution of intrabranch transitions for $\lambda =0$
branch is 
\begin{eqnarray}
\sigma _{xy,intra,0}^{zz} &=&-\frac{q}{\pi \hbar }\frac{\langle \lambda
n_{\lambda }|\hat{k}_{x}|\lambda n_{\lambda }-1\rangle \langle \lambda
n_{\lambda }-1|\hat{J}_{2,x}^{zz}|\lambda n_{\lambda }\rangle }{\omega
_{n_{\lambda }}-\omega _{n_{\lambda }-1}}  \notag \\
&=&-\frac{q}{\pi \hbar }\frac{1}{\omega _{n_{\lambda }}-\omega _{n_{\lambda
}-1}}\frac{\hbar ^{2}(n_{\lambda }-1)}{6m^{\ast }l^{2}}  \notag \\
&=&-\frac{q\hbar (l^{2}k_{0}^{2}-2)}{12m^{\ast }\pi l^{2}\omega _{c}}.
\end{eqnarray}%
Finally we get the rank-2 spin-tensor Hall conductivity from the contribution
of intrabranch transitions by adding the results for $\lambda =\pm 1,0$ as
follows, 
\begin{eqnarray}
\sigma _{xy,intra}^{zz} &=&\sigma _{xy,intra,\pm 1}^{zz}+\sigma
_{xy,intra,0}^{zz}  \notag \\
&=&-\frac{q}{8\pi }+\frac{q}{12\pi }\gamma ^{2}.
\end{eqnarray}%

In summary, the rank-2 STH conductivity is 
\begin{eqnarray}
\sigma _{xy}^{zz} &=&\sigma _{xy,inter}^{zz}+\sigma _{xy,intra}^{zz}  \notag
\\
&=&\frac{q}{12\pi }\gamma ^{2}=\frac{m^{\ast 2}\alpha ^{2}c}{6\pi \hbar ^{3}B%
}.
\end{eqnarray}%
We note that in the above calculations, only two adjacent states near the
Fermi surface, one below the Fermi surface and the other above the Fermi
surface, contribute to the the rank-2 STH conductivity. The reason is as
follows. We take the intrabranch transition between two non-adjacent states,
one state below the Fermi surface defined as $\left\vert \lambda n_{\lambda
}-1\right\rangle $ and the other state above the Fermi surface denoted by $%
\left\vert \lambda n_{\lambda }+1\right\rangle $, as an example. Due to $%
\langle \lambda n_{\lambda }+1|\hat{k}_{x}|\lambda n_{\lambda }-1\rangle =0$%
, the $\sigma _{xy,intra,\pm 1}^{zz}=\sigma _{xy,intra,0}^{zz}=0$.
Obviously, the STH conductivity from the contribution of other intrabranch
transitions between non-adjacent states takes the same result. Therefore,
the intrabranch transitions between non-adjacent states do not contribute to
STH conductivity.

\subsection{Rank-1 spin Hall conductivity}

For the given system, the rank-1 spin current operator can be written as $%
\hat{J}_{x}^{z}=\frac{1}{2}\left\{ \hbar \hat{F}_{z},\hat{v}_{x}\right\} $,
which leads to
\begin{eqnarray}
\hat{J}_{x}^{z}=\frac{\hbar ^{2}\hat{k}_{x}}{m^{\ast }}\hat{F}_{z}+\frac{\alpha }{2%
\sqrt{2}}(\hat{\lambda} _{2}^{\ast }+\hat{\lambda} _{7}^{\ast }).
\end{eqnarray}%
The calculation is similar to that in the rank-2 STH conductivity, which also
consists of the contribution of intrabranch and interbranch transitions.
First, the rank-1 spin Hall conductivity from the contribution of the
interbranch transitions is
\begin{eqnarray}
\sigma _{xy,inter}^{zz} &=&-\frac{q}{\pi \hbar }\sum_{\lambda n\lambda
^{\prime }n^{\prime }}\frac{\langle \lambda n|\hat{k}_{x}|\lambda ^{\prime
}n^{\prime }\rangle \langle \lambda ^{\prime }n^{\prime }|\hat{J}%
_{x}^{z}|\lambda n\rangle }{\omega _{\lambda n}-\omega _{\lambda ^{\prime
}n^{\prime }}}  \notag \\
&=&0,
\end{eqnarray}%
and the rank-1 spin Hall conductivity from the contribution of the
intrabranch transitions is
\begin{eqnarray}
\sigma _{xy,intra}^{z} &=&-\frac{q}{\pi \hbar }\frac{\langle \lambda
n_{\lambda }|\hat{k}_{x}|\lambda n_{\lambda }-1\rangle \langle \lambda
n_{\lambda }-1|\hat{J}_{x}^{z}|\lambda n_{\lambda }\rangle }{\omega
_{n_{\lambda }}-\omega _{n_{\lambda }-1}}  \notag \\
&=&0,
\end{eqnarray}%
where $\lambda =-1,0,1$. Therefore, the rank-1 spin Hall conductivity is $%
\sigma _{xy}^{z}=\sigma _{xy,inter}^{z}+\sigma _{xy,intra}^{z}=0$.

\subsection{Rank-0 Hall conductivity}

By the definition $\hat{J}_{0}=-q\hat{v}_{x}$, charge current operator is
written as
\begin{eqnarray}
\hat{J}_{0}=-\frac{q\hbar \hat{k}_{x}}{m^{\ast }}+\frac{q\alpha }{\sqrt{2}%
\hbar }(\hat{\lambda}_{2}+\hat{\lambda}_{7}^{\ast }).
\end{eqnarray}%
The calculation of rank-0 Hall conductivity is also similar to rank-2 STH
conductivity. The contribution of interbranch and intrabranch transitions are
respectively given by 
\begin{eqnarray}
\sigma _{xy,inter}^{zz} &=&-\frac{q}{\pi \hbar }\sum_{\lambda n\lambda
^{\prime }n^{\prime }}\frac{\langle \lambda n|\hat{k}_{x}|\lambda ^{\prime
}n^{\prime }\rangle \langle \lambda ^{\prime }n^{\prime }|\hat{J}%
_{0}|\lambda n\rangle }{\omega _{\lambda n}-\omega _{\lambda ^{\prime
}n^{\prime }}}  \notag \\
&=&0,  \label{B15} \\
\sigma _{xy,intra} &=&-\frac{q}{\pi \hbar }\frac{\langle \lambda n_{\lambda
}|\hat{k}_{x}|\lambda n_{\lambda }-1\rangle \langle \lambda n_{\lambda }-1|%
\hat{J}_{0}|\lambda n_{\lambda }\rangle }{\omega _{n_{\lambda }}-\omega
_{n_{\lambda }-1}}  \notag \\
&=&\frac{q^{2}(-3+2\gamma ^{2}+6\eta )}{4\pi \hbar }.  \label{B16}
\end{eqnarray}%



In summary, the rank-0 Hall conductivity is
\begin{eqnarray}
\sigma_{xy}&=& \sigma_{xy,inter}+\sigma_{xy,intra}  \notag \\
&=&\frac{q^{2}(-3+2\gamma^{2}+6\eta)}{4\pi\hbar}  \notag \\
&=& \frac{q(-3\hbar^{3}qB+4m^{*2} \alpha^{2}c+6m^{*}\hbar^{3}c\omega_{f})}{%
4\pi\hbar^{4}B},
\end{eqnarray}
as shown by the solid curve in Fig. \ref{Fig3}. Since there are some
approximations in the above analytical calculation, we numerically compute
rank-0 Hall conductance by Eqs.(\ref{B15}) and (\ref{B16}) directly, as
shown by the blue dots in Fig. \ref{Fig3}. The analytical results and
numerical results show good alignments. 

\begin{figure}[t]
\centering
\includegraphics[width=0.35\textwidth]{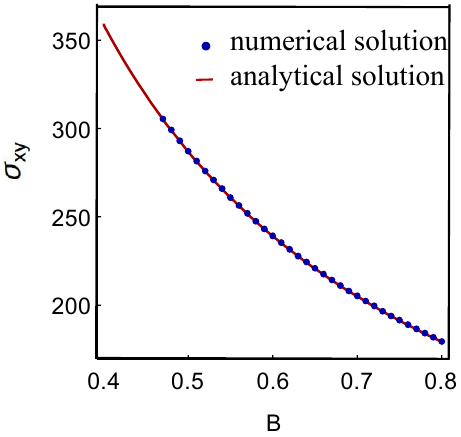}
\caption{ Numerical and analytical results for the rank-0 Hall conductivity
versus the magnetic field, $m^{\ast }=\protect\alpha=c=1$. }
\label{Fig3}
\end{figure}

\section{Rank-0 Hall conductivity in a spin-1/2 system}

\label{notation2} The Hamiltonian for 2D electron gas with Rashba spin-orbit
coupling subjected to an external perpendicular magnetic field is given by
\begin{eqnarray}
\hat{H}=\frac{\hbar ^{2}\bm{\hat{k}}^{2}}{2m^{\ast }}+\alpha (%
\bm{\hat{\sigma}}\times \bm{\hat{k}})\cdot \bm{z}.
\end{eqnarray}%
The charge current operator is given by
\begin{eqnarray}
\hat{v}_{x} &=&\partial {\hat{H}}/\partial {\hat{p}_{x}}=\frac{\hbar \hat{k}%
_{x}}{m}-\alpha \hat{\sigma}_{y}, \\
\hat{J}_{0} &=&-q\hat{v}_{x}=\frac{-q\hbar \hat{k}_{x}}{m}+q\alpha \hat{%
\sigma}_{y}.
\end{eqnarray}%
According to the Kubo formula, the rank-0 Hall conductivity from the
contribution of interbranch and intrabranch transitions is
\begin{eqnarray}
\sigma _{xy,inter} &=&0, \\
\sigma _{xy,intra} &=&\frac{q^{2}(-1+\gamma ^{2}+2\eta )}{2\pi \hbar }.
\end{eqnarray}

In summary, the rank-0 Hall conductivity is $\sigma_{xy}=\sigma_{xy,inter}+%
\sigma_{xy,intra}=\frac{q^{2}(-1+\gamma^{2}+2\eta)}{2\pi\hbar}$. Therefore,
it is not a universal value in a spin-1/2 system.

\section{Rank-2 STH conductivity under the pseudospin basis}

\label{notation3} Pseudospin dynamics analysis can give more intuitions. We
first define pseudospin by
\begin{equation}
\left\vert \Uparrow \right\rangle =\frac{1}{\sqrt{2}}(\left\vert \tilde{%
\uparrow}\right\rangle +\left\vert \tilde{\downarrow}\right\rangle ), \\
\left\vert \Downarrow \right\rangle =\left\vert \tilde{\nmid}\right\rangle ,
\notag \\
\left\vert \o \right\rangle =\frac{1}{\sqrt{2}}(\left\vert \tilde{\uparrow}%
\right\rangle -\left\vert \tilde{\downarrow}\right\rangle ),  \notag
\end{equation}%
where $\left\vert \Uparrow \right\rangle $ and $\left\vert \Downarrow
\right\rangle $ are orthogonal to $\left\vert \o \right\rangle $.
When projected onto the pseudospin subspace $\{\left\vert \Uparrow
\right\rangle ,\left\vert \Downarrow \right\rangle \}$, $\mathcal{\hat{H}}%
_{Z}$ yields
\begin{equation}
\hat{H}_{s}=\hat{P}_{s}\mathcal{\hat{H}}_{Z}\hat{P}_{s}^{-1}=\frac{%
\bm{\hat{p}}^{2}}{2m^{\ast }}-\frac{\alpha }{\hbar }\bm{\hat{\sigma}}\cdot (%
\bm{z}\times \bm{\hat{p}}),
\end{equation}%
where the spin operator $\bm{\hat{\sigma}}$ acts on the pseudospin, and $%
\hat{P}_{s}=\left\vert \Uparrow \right\rangle \left\langle \Uparrow
\right\vert +\left\vert \Downarrow \right\rangle \left\langle \Downarrow
\right\vert $ is the projection operator. Note that, when projected into the
pseudospin subspace, the spin tensor $N_{zz}$ changes to
\begin{equation}
\hat{P}_{s}\hat{N}_{zz}\hat{P}_{s}^{-1}=\frac{1}{2}\hat{\sigma}_{z}-\frac{1}{%
6}\hat{\sigma}_{0},
\end{equation}%
where $\hat{\sigma}_{0}$ is an identity operator. The rank-2 spin-tensor
current then becomes
\begin{equation}
\hat{J}_{2,x}^{zz}=\frac{1}{2}\{\frac{\hbar }{2}\hat{\sigma}_{z},\hat{v}%
_{x}\}-\frac{1}{6}\hat{\sigma}_{0}\hat{v}_{x}. \label{D3}
\end{equation}

In the pseudospin subspace $\{\left\vert \Uparrow \right\rangle ,\left\vert
\Downarrow \right\rangle \}$, the contributions of the interbranch
transitions and interbranch transitions to the conductivity are respectively
given by
\begin{eqnarray}
\notag
\sigma _{xy,inter}^{zz} &=&-\frac{q}{\pi \hbar }\sum_{\lambda n\lambda
^{\prime }n^{\prime }}\frac{\langle \lambda n|\hat{k}_{x}|\lambda ^{\prime
}n^{\prime }\rangle \langle \lambda ^{\prime }n^{\prime }|\hat{J}%
_{2,x}^{zz}|\lambda n\rangle }{\omega _{\lambda n}-\omega _{\lambda ^{\prime
}n^{\prime }}}   \\
&=&\frac{q}{8\pi }, \\
\sigma _{xy,1,intra}^{zz} &=&-\frac{q}{\pi \hbar }\frac{\langle \lambda
n_{\lambda }|\hat{k}_{x}|\lambda n_{\lambda }-1\rangle \langle \lambda
n_{\lambda }-1|\hat{J}_{2,x}^{zz}|\lambda n_{\lambda }\rangle }{\omega
_{n_{\lambda }}-\omega _{n_{\lambda }-1}}  \notag \\
&=&\frac{q\eta }{6\pi }+\frac{q(-5+4\gamma ^{2})}{24\pi }+\frac{q(3-4\gamma
^{2}+\gamma ^{4})}{96\pi \gamma ^{2}\eta }  \notag \\
&-&\frac{q\gamma ^{2}}{12\pi }.
\end{eqnarray}%
Thus, the STH conductivity in the subspace $\{\left\vert \Uparrow
\right\rangle ,\left\vert \Downarrow \right\rangle \}$ is given by
\begin{eqnarray}
\sigma _{xy,1}^{zz} &=&\sigma _{xy,1,inter}^{zz}+\sigma _{xy,1,intra}^{zz}
\notag \\
&=&\frac{q\eta }{6\pi }+\frac{q\gamma ^{2}}{12\pi }-\frac{q}{12\pi }.
\end{eqnarray}

We would like to remark that the contributions to spin Hall conductivity from
intrabranch and interbranch transitions cancel out in the spin-1/2 system
with Rashba spin-orbit coupling \cite{21}. However, in this work while the
projected Hamiltonian $\hat{H}_{Z}$ takes the same form as that in reference
\cite{21}, the projected spin tensor $\hat{N}_{zz}$ has an extra term
proportional to $\hat{\sigma}_{0}$ besides the term proportional to $\hat{%
\sigma}_{z}$, leading to $\frac{1}{6}\hat{\sigma}_{0}\hat{v}_{x}$ in $\hat{J}%
_{2,x}^{zz}$ as in Eq.(\ref{D3}). This gives rise to that the contributions
to STH conductivity from intrabranch and interbranch transitions can not
cancel out.

If $\mathcal{\hat{H}}_{Z}$ is projected to the space spaned by $\{\left\vert
\o \right\rangle \}$, we will have the following contribution to the STH
conductivity
\begin{eqnarray}
\sigma _{xy,2}^{zz}=-\frac{q(2\eta -1)}{12\pi }.
\end{eqnarray}

In summary, the total STH conductivity from the contribution of transitions
in the subspace $\{\left\vert \Uparrow \right\rangle ,\left\vert \Downarrow
\right\rangle \}$ and \{$\left\vert \o \right\rangle \}$ is
\begin{equation}
\sigma _{xy}^{zz}=\sigma _{xy,1}^{zz}+\sigma _{xy,2}^{zz}=\frac{q\gamma ^{2}%
}{12\pi }.
\end{equation}%
We obtained the same result that have been obtained in Section 1 of Appendix %
\ref{notation1}.

\section{Rank-2 STH conductivity by perturbation theory}
\label{notation4}}
Under the weak magnetic field, we treat the vector potential $\bm{A}=xB\hat{y}$ as a perturbation. The Hamiltonian $\mathcal{\hat{H}}$ up to the first order in $\bm{A}$ can be rewritten as
\begin{eqnarray}
\mathcal{\hat{H}} = H_{0}+\frac{q p_{y}}{c m^*}xB+\frac{1}{\sqrt{2}}\frac{%
\alpha q}{\hbar c}xB(\bm{\hat{\tau}}_{T,x}+\bm{\hat{\tau}}_{V,x}^*). \label{E1}
\end{eqnarray}
Diagonalizing the Hamiltonian in Eq.(\ref{E1}) yields three eigenvalues,
\begin{eqnarray}
    E_{1}&=&\frac{\hbar^{2}k^{2}}{2m^{*}}-\alpha k+i p_{y}\Gamma-\frac{\alpha m^{*}e^{i\theta}}{\hbar}\Gamma +\Lambda,
\end{eqnarray}
\begin{eqnarray}
E_{2}&=&\frac{\hbar^{2}k^{2}}{2m^{*}},\\
    E_{3}&=&\frac{\hbar^{2}k^{2}}{2m^{*}}+\alpha k+i p_{y}\Gamma+\frac{\alpha m^{*}e^{i\theta}}{\hbar}\Gamma +\Lambda,
    \end{eqnarray}
with the corresponding vectors
\begin{eqnarray}
    \vert1\rangle&=&\vert \overline{1}k\rangle+\frac{i p_{y}}{2\alpha k}\Gamma\vert1k\rangle+\frac{m^{*}e^{i\theta}}{2\hbar k}\Gamma\vert1k\rangle ,\\
    \vert2\rangle&=&\vert0k\rangle , \\
    \vert3\rangle&=&\vert1k\rangle-\frac{i p_{y}}{2\alpha k}\Gamma\vert\overline{1}k\rangle+\frac{m^{*}e^{i\theta}}{2\hbar k}\Gamma\vert\overline{1}k\rangle.
\end{eqnarray}%
where $\Gamma=\frac{\omega_{c} (k e^{-i\theta}-k_{x})}{2k^{2}}$, $\Lambda=(\frac{m^{\ast 2} \alpha e^{2i\theta}}{2\hbar^{2} k}+\frac{p_{y}^{2}}{2\alpha k})\Gamma^{2}$.

The velocity operator and the current operator are respectively given by
\begin{eqnarray}
J_{2,x}^{zz}&=&\frac{\hbar^{2}k_{x}}{m} N_{zz}+\frac{\alpha}{2\sqrt{2}}(\lambda_{2}-\lambda_{7}),\\
v_{y}&=&\frac{\hbar k_{y}}{m}+\frac{\alpha}{\sqrt{2}\hbar}(\lambda_{1}+\lambda_{6}).
\end{eqnarray}
By using Kubo formula as in Eq.(\ref{sigmazz}), we obtain the same rank-2 STH conductivity $\sigma _{xy,inter}^{zz}=\frac{q}{8\pi }$ as that by Landau level construction in Sec. \ref{2}. Here, we also note that only the interbranch transitions  between the highest branch ($\lambda=1$) and the lowest branch ($%
\lambda=-1$) contribute to the rank-2 STH conductivity.
However, we cannot derive the contribution to the rank-2 STH conductivity from intrabranch transitions since the energy levels are continuous by perturbation method and it is impossible to distinguish states above and below the Fermi surface. This is why Landau levels are introduced in this work and a previous work by Rashba   \cite{21}, where one can distinguish states above or below the Fermi surface when the Fermi energy is at the gap between Landau levels.

\end{document}